# Interdependent Relationships in Game Theory: A Generalized Model

Jiawei Li

School of Natural Sciences, University of Stirling, UK, FK9 4LA

Email: lij@cs.stir.ac.uk

**Abstract:** A generalized model of games is proposed, in which cooperative games and non-cooperative games are special cases. Some games that are neither cooperative nor non-cooperative can be expressed and analyzed. The model is based on relationships and supposed relationships between players. A relationship is a numerical value that denotes how one player cares for the payoffs of another player, while a supposed relationship is another numerical value that denotes a player's belief about the relationship between two players. The players choose their strategies by taking into consideration not only the material payoffs but also relationships and their change. Two games, a prisoner's dilemma and a repeated ultimatum game, are analyzed as examples of application of this model.

**Keywords**: game theory, cooperative games, non-cooperative games, relationship.

## 1. Introduction

There have been two kinds of researches in game theory: cooperative games theory pioneered by Von Neumann and Morgenstern (1953) and non-cooperative games theory developed by Nash (1951). When analyzing a game, one first needs to confirm what type the game is because absolutely different methods will be used for two types of games. Methods for non-cooperative games are based on Nash equilibrium, various perfects of Nash equilibrium (e.g., *Strong Nash Equilibrium* by Aumann (1959), *Subgame Perfect Nash Equilibrium* and *Trembling Hand Perfect Equilibrium* by Selten (1965, 1975), *Bayesian Nash Equilibrium* and *Strict Nash Equilibrium* by Harsanyi (1967, 1973)) and the folk theorems, while cooperative games are analyzed by means of coalitions, core and Shapley value (Shapley (1953)). The most important work comes from non-cooperative analysis, although some scholars regain interest in cooperative games recently.

A question is whether or not any games can be just categorized into these two types? Consider contract bridge, the card game that is played by four players in two competing partnerships. This game is neither cooperative nor non-cooperative because the relations between individual players are not identical. There are both cooperation between partners and competition between two partnerships.

The players in a non-cooperative game only care for their own payoffs whilst the players in a cooperative game care for the payoffs of other players in the coalition as equally important as their own payoffs. If we use a numerical value to denote how much one player cares for another player's payoff, this value will be zero for non-cooperative games and one for cooperative games. One may naturally ask what if this value is set to be neither zero nor one.

In this paper, we introduce the concepts of relationship and supposed relationship. A relationship is a numerical value denoting how much one player cares for another player's payoff. A supposed relationship is a numerical value denoting a player's belief about how much one player cares for another player's payoff. Relationships and supposed relationships are determined by the players, and they are changeable in different stages of games. We propose a relationship model of games, in which strategic interaction among players is determined by the material payoffs, relationships and the players' beliefs about relationships. Cooperative games and non-cooperative games, as well as those games that are neither cooperative nor non-cooperative can be expressed and analyzed by using this model.

Interpersonal relationship has not attracted interests of research in game theory although it has long been an important topic of research in many social science disciplines such as psychology and politics (Kelley (2013), Heider (2013)). Game theorists take it for granted that the relationships between players are predetermined and they will never change during the strategic interactions of players.

Interdependent preference, which denotes that a player's preference depends on his/her opponent's payoff as well as his/her own payoff, has been used to explain cooperation phenomena in experimental economics (e.g. Bolton 1991, 2000, Binmore et al. 2002, Ochs and Roth 1989, Samuelson 2001). It is quite similar to the relationship concept defined in this paper. Reputation effect pioneered by Kreps and Wilson (1982) and Milgrom and Roberts (1982) has been introduced into game theory to explain cooperation in repeated non-cooperative games. Relationship is obviously different from reputation in that reputation is independent of the game model and then has less effect in one-shot games. Cooper et al. (1996) and Chan (2000) suggested that reputation was unnecessarily the unique factor leading to cooperation in either infinitely or finitely repeated games.

In the rest of the paper, we introduce a novel game model taking into consideration the relationship and relationship change among players. We also show the application of our model in a large set of games that are neither cooperative nor non-cooperative.

## 2. A Relationship Model

*Definition* 1: For players *i* and *j* in a game, a ***relationship*** $R_{ij}$ is a numerical value denoting how much player *i* cares for player *j*'s payoff. Specially, there is $R_{ii}=1$.

Player *i*'s attitude toward *j* is non-cooperative when $R_{ij}=0$, and cooperative when $R_{ij}=1$. We call it a sub-cooperative attitude when $0<R_{ij}<1$, a hostile attitude when $R_{ij}<0$, and a dedicated attitude when $R_{ij}>1$. Sub-cooperative attitude is common in interpersonal relationships. People do not assign equal weights to the payoffs of their own and that of others although they do care for the payoffs of others. One typical example of hostile attitude can be found in suicide terrorist attacks. The terrorists must have a hostile attitude towards the targeted government so that they would rather give up their lives in the attacks. Dedicated attitude can be found in special real world instances, for example, a parent's attitude toward their children, a religious believer's attitude toward their god.

*Definition* 2: For players *i*, *j* and *k* in a game, a ***supposed relationship*** ${}_k\bar{R}_{ij}$ is a numerical value denoting how much player *k* thinks player *i* cares for player *j*'s payoff. Specially, there is ${}_k\bar{R}_{ii}=1$.

Player *k* thinks that player *i*'s attitude toward *j* is non-cooperative when ${}_k\bar{R}_{ij}=0$, cooperative when ${}_k\bar{R}_{ij}=1$, sub-cooperative when $0<{}_k\bar{R}_{ij}<1$, hostile when ${}_k\bar{R}_{ij}<0$, or dedicated when ${}_k\bar{R}_{ij}>1$. In a game with complete information, there are $R_{ij}={}_k\bar{R}_{ij}$, which means that the relationships between players are known to all players. However, in a game with incomplete information, a supposed relationship is not necessarily equivalent to the corresponding relationship.

Obviously cooperative and non-cooperative games are the special cases in the set of games defined by different values of $[R_{ij}]$ and $[_k\bar{R}_{ij}]$. For example, a game is non-cooperative when $R_{ij} =\, _k\bar{R}_{ij} = 0$ for any $i \neq j$ and k, while a game is cooperative when $R_{ij} =\, _k\bar{R}_{ij} = 1$ for any $i, j$ and $k$.

Note that $\{R_{ij}\}$ can be considered as a subset of $\{_k\bar{R}_{ij}\}$ because every player knows his/her own relationship and thus there must be $_i\bar{R}_{ij} = R_{ij}$. We keep both variables in this paper to distinguish a relationship from other player's belief about it.

*Definition* 3: The relationship model of an *n*-player game $G$ is a 4-tuple, $G = \{I, S, U, \bar{R}\}$, where $I = \{1, \cdots, n\}$ is the player set, $S = \{S_1, \cdots, S_n\}$ is the strategy set, $U$ is the payoff set, and $\bar{R}$ is the supposed relationship set.

*Definition* 4: A supposed payoff of player $i$, $_i\bar{u}_j$, denotes how much player $i$ thinks player $j$'s payoff is by taking player $i$'s supposed relationships into consideration.

Given a strategy profile $(s_1, \cdots, s_n)$, player $j$'s payoff $u_j(s_1, \cdots, s_n)$, and player $i$'s supposed relationships $\{_i\bar{R}\}$, the supposed payoff, $_i\bar{u}_j$, is computed by

$$\forall i,\ _i\bar{u}_j(s_1, \cdots, s_n) = \begin{cases} \sum_{k=1}^{n} R_{ik} u_k & \text{if } j = i \\ \sum_{k=1}^{n}\, _i\bar{R}_{jk} u_k & \text{if } j \neq i \end{cases} \quad (1)$$

Under the relationship model, the players in a game choose their strategies according to their supposed payoffs. For example, player $i$'s choice is determined by $i$'s supposed payoffs, $[_i\bar{u}]$. We prove in the following theorem that there must exist at least one Nash equilibrium for every game when players make choices according to their supposed payoffs.

**Theorem 1**: Under relationship model, there must exist a Nash equilibrium for every game.

Proof: Consider an arbitrary player $i$ in an *n*-player game ($i \in N$). Let $[_i\bar{u}]$ denote $i$'s supposed payoffs. Since $[_i\bar{u}]$ is a complete payoff matrix, there must be a Nash equilibrium strategy profile for $[_i\bar{u}]$, as Nash proved in [4]. Let $s_i$ be player $i$'s strategy in this profile. The strategies of all players form a new strategy profile $\{s_i\}, i = \{1, \cdots, n\}$. It is obvious that $\{s_i\}$ is a Nash equilibrium because every player $i$ has no incentive to deviate from $s_i$. □

Let's analyze the prisoner's dilemma as an example to show how to use the relationship model of games. The payoff matrix of a prisoner's dilemma is shown in Fig.1.

|  | | y Player | |
|---|---|---|---|
|  |  | C | D |
| x Player | C | 3, 3 | 0, 5 |
|  | D | 5, 0 | 1, 1 |

Figure 1 Two players choose between Cooperate (*C*) and Defect (*D*) in the prisoner's dilemma. The numbers in each cell denote the payoffs of players *x* and *y* respectively.

Let $R_{xy}$ and $R_{yx}$ denote the relationships between the players *x* and *y*, and $_y\bar{R}_{xy}$ and $_x\bar{R}_{yx}$ the corresponding supposed relationships of *y* and *x*. The supposed payoffs of player *x* can be computed according to (1). They are expressed as a matrix as shown in Fig. 2.

|  | | y Player | |
|---|---|---|---|
|  |  | C | D |
| x Player | C | 3+3 $R_{xy}$, 3+3 $_x\bar{R}_{yx}$ | 5 $R_{xy}$, 5 |
|  | D | 5, 5 $_x\bar{R}_{yx}$ | 1+ $R_{xy}$, 1+ $_x\bar{R}_{yx}$ |

Figure 2 The supposed payoffs of player *x*.

What player *x* chooses between *C* and *D* depends on the values of $R_{xy}$ and $_x\bar{R}_{yx}$. Choosing *C* is the dominant for player *x* when there are both $3+3R_{xy} \geq 5$ and $5R_{xy} \geq 1+R_{xy}$, or $R_{xy} \geq \frac{2}{3}$. Strategy *D* is dominant when there are both $3+3R_{xy} \leq 5$ and $5R_{xy} \leq 1+R_{xy}$, or $R_{xy} \leq \frac{1}{4}$. In the case of $\frac{1}{4} < R_{xy} < \frac{2}{3}$, *C* is dominant when $_x\bar{R}_{yx} \leq \frac{1}{4}$ and *D* is dominant when $_x\bar{R}_{yx} \geq \frac{2}{3}$. No strategy is dominant for player *x* when there are $\frac{1}{4} < R_{xy}, _x\bar{R}_{yx} < \frac{2}{3}$.

Similarly, the supposed payoffs of player *y* are computed and expressed as a matrix as shown in Fig. 3,

|  | | y Player | |
|---|---|---|---|
|  |  | C | D |
| x Player | C | 3+3 $_y\bar{R}_{xy}$, 3+3 $R_{yx}$ | 5 $_y\bar{R}_{xy}$, 5 |
|  | D | 5, 5 $R_{yx}$ | 1+ $_y\bar{R}_{xy}$, 1+ $R_{yx}$ |

Figure 3 The supposed payoffs of player *y*.

Comparing Fig. 3 with Fig. 2, the supposed payoffs of two players will be identical if $_y\bar{R}_{xy}$ and $R_{yx}$ are replaced by $R_{xy}$ and $_x\bar{R}_{yx}$. *C* is the dominant strategy for player *y* when $R_{yx} \geq \frac{2}{3}$, while *D* is dominant when $R_{yx} \leq \frac{1}{4}$. In the case of $\frac{1}{4} < R_{yx} < \frac{2}{3}$, *C* is dominant when $_y\bar{R}_{xy} \leq \frac{1}{4}$ while *D* is dominant when $_y\bar{R}_{xy} \geq \frac{2}{3}$. No strategy is dominant when there are $\frac{1}{4} < R_{yx}, _y\bar{R}_{xy} < \frac{2}{3}$.

If we consider the prisoner's dilemma as a game of complete information, there should be $R_{xy} = {_y\bar{R}_{xy}}$ and $R_{yx} = {_x\bar{R}_{yx}}$, which means that the relationships are known to both players. The supposed payoffs of different players are identical in a game of complete information. The players' choices depend on the exact values of $R_{xy}$ and $R_{yx}$. For example, the strategy profile (C, C) is dominant when $R_{xy} \geq \frac{2}{3}$ and $R_{yx} \geq \frac{2}{3}$; (D, D) is dominant when $R_{xy} \leq \frac{1}{4}$ and $R_{yx} \leq \frac{1}{4}$; (C, D) is dominant when $R_{xy} \geq \frac{2}{3}$ and $R_{yx} \leq \frac{1}{4}$; and (D, C) is dominant when $R_{xy} \leq \frac{1}{4}$ and $R_{yx} \geq \frac{2}{3}$. The game has multiple equilibria and no strategy profile is dominant when there are $\frac{1}{4} < R_{yx}, R_{xy} < \frac{2}{3}$.

The strategy interaction in a game of incomplete information can be much diverse and complex because the relationships of one player are unknown to others. The supposed relationships are not necessarily equivalent to the corresponding relationship in a game of incomplete information. Since the supposed relationships are private information, the players would take advantage of them in strategic interactions. For example, player $i$ attempts to exploit the opponent by setting $R_{ij} <_i \overline{R}_{ji}$. On the other hand, $R_{ij} >_i \overline{R}_{ji}$ reflects an altruistic attitude of player $i$ toward $j$. There should be $R_{ij} =_i \overline{R}_{ji}$ if player $i$ adopts a tit-for-tat strategy, or in other words, player $i$ wants to treat the opponent exactly same as what the opponent treats themselves.

If we consider the prisoner's dilemma as a game of incomplete information, equilibrium analysis can be made given the values of $R_{xy}$, $R_{yx}$, $_y\overline{R}_{xy}$, and $_x\overline{R}_{yx}$. Suppose that there are $R_{xy} = R_{yx} = \frac{1}{3}$ and $_y\overline{R}_{xy} = {_x\overline{R}_{yx}} = \frac{1}{5}$, the supposed payoffs can be computed as shown in Fig. 4. From Fig. 4(a), D is dominant strategy for player $y$. Player $x$ will choose C given that he/she believes that the other player will choose D. Similarly, from Fig. 4(b), player $y$ will also choose C given that he/she thinks the opponent would choose D. Thus, strategy profile (C, C) will be the outcome. In this scenario, both players have an altruistic attitude towards another player and thus they would choose C in order to maximize the other player's payoff.

|  |  | y Player | |
|---|---|---|---|
|  |  | C | D |
| x Player | C | 4, $3\frac{3}{5}$ | $1\frac{2}{3}$, 5 |
|  | D | 5, 1 | $1\frac{1}{3}$, $1\frac{1}{5}$ |

(a)

|  |  | y Player | |
|---|---|---|---|
|  |  | C | D |
| x Player | C | $3\frac{3}{5}$, 4 | 1, 5 |
|  | D | 5, $1\frac{2}{3}$ | $1\frac{1}{5}$, $1\frac{1}{3}$ |

(b)

Figure 4 (a) Supposed payoffs of player $x$. (b) Supposed payoffs of player $y$.

## 3. Repeated games

A repeated game $G^T$ is a 4-tuple, $G = \{I, S, U, \overline{R}^T\}$, where $T$ is the number of iteration. The players in a repeated game will have to take relationship change into consideration when choosing their strategies. Relationship change reflects the complexity of intelligent decision making. It could be a complex action depending on how the players retrieve information from previous interactions with other players and how they update their supposed relationships.

One reason for relationship change in repeated games lies in the fact that previous strategic interactions provide new information about relationships so that the players should update their supposed relationships. Take the iterated prisoner's dilemma as an example. Suppose that two players play the prisoner's dilemma with payoff matrix as shown in Fig. 1 repeatedly, and there are $R_{xy} = R_{yx} = \frac{1}{3}$ and $_y\overline{R}_{xy} = {_x\overline{R}_{yx}} = \frac{1}{5}$ at the beginning of game. According to two players' supposed payoffs shown in Fig. 4, they choose (C, C) in the first round. After playing the first round, two players realize that they have underestimated $_y\overline{R}_{xy}$ and $_x\overline{R}_{yx}$ because the other player should have chosen D instead of C according to their original evaluations of $_y\overline{R}_{xy}$ and $_x\overline{R}_{yx}$. Two players should then increase their supposed relationships to some values greater than $\frac{1}{4}$.

Relationship change may take place in games of complete information as well. Assume that there are $R_{xy} \neq R_{yx}$ in the above example. A tit-for-tat player will update his/her relationship to make sure $R_{xy} = R_{yx}$.

In the following example we analyze a repeated ultimatum game in order to show how players take into consideration relationship change in game playing.

Let's consider an infinitely repeated ultimatum game. In each round, Row player proposes an offer of dividing one dollar between two players. If Column player accepts the offer, they receive the corresponding share. Otherwise, both players receive nothing. The minimum division of one dollar is one cent. Fig. 5 shows the payoff matrix of this game. The supposed payoffs of Column player are shown in Fig. 6.

|  |  | *Column* player | |
|---|---|---|---|
|  |  | *Accept* | *Reject* |
|  | *0* | 1, 0 | 0, 0 |
|  | *0.01* | 0.99, 0.01 | 0, 0 |
| Row player | ... | ... | 0, 0 |
|  | *α* | 1-α, α | 0, 0 |
|  | ... | ... | 0, 0 |
|  | *0.99* | 0.01, 0.99 | 0, 0 |

Figure 5 An ultimatum game in which Row player makes an offer of dividing a dollar to Column player.

|  |  | *Column* player | |
|---|---|---|---|
|  |  | *Accept* | *Reject* |
|  | *0* | 1, 0 | 0, 0 |
| Row player | ... | ... | 0, 0 |
|  | *α* | $1 - \alpha + {_c\overline{R}_{RC}}\alpha$, $\alpha + (1 - \alpha) R_{CR}$ | 0, 0 |
|  | ... | ... | 0, 0 |

Figure 6 The supposed payoffs of Column player.

The supposed payoffs of two players can be computed according to the relationships and supposed relationships of $R_{RC}, R_{CR}, {_R\overline{R}_{CR}}$, and ${_c\overline{R}_{RC}}$. For an offer $\alpha$, Row player's supposed payoff profile is ($1 - \alpha + R_{RC}\alpha$, $\alpha + (1 - \alpha) {_R\overline{R}_{CR}}$) while Column player's supposed payoff profile is ($1 - \alpha + {_c\overline{R}_{RC}}\alpha$, $\alpha + (1 - \alpha) R_{CR}$).

In order for Column player to accept $\alpha$, there must be $\alpha + (1 - \alpha) R_{CR} > 0$, or $\alpha > \frac{-R_{CR}}{1 - R_{CR}}$. On the other hand, for Row player there must be $1 - \alpha + R_{RC}\alpha > 0$, or $\frac{1}{1 - R_{RC}} > \alpha$. Taking into consideration the supposed relationships of two players, Row player will offer $\alpha$ satisfying $\frac{1}{1-R_{RC}} > \alpha > \frac{-{}_R\overline{R}_{CR}}{1-{}_R\overline{R}_{CR}}$ and Column player will accept $\alpha$ satisfying $\frac{1}{1-{}_C\overline{R}_{RC}} > \alpha > \frac{-R_{CR}}{1-R_{CR}}$.

When ${}_R\overline{R}_{CR} \geq 0$, Row player thinks Column player would accept any offer $\alpha > 0$ so Row player would offer $\alpha = 0.01$ to maximize the payoff. This is what non-cooperative game theory has predicted. However, Column player could choose $R_{CR} < 0$ and would definitely reject the offer of $\alpha = 0.01$. In an infinitely repeated game, two players will reach an agreement with the offer in range of $\frac{1}{1-R_{RC}} > \alpha > \frac{-R_{CR}}{1-R_{CR}}$ even if two players cannot communicate with each other. Consider the case of $R_{RC} = R_{CR} = {}_R\overline{R}_{CR} = {}_C\overline{R}_{RC} = -0.5$. Column player would reject any offer with $\alpha < \frac{1}{3}$, which transferred the information about Column player's relationship value to Row player. Row player would then have to increase the offer in order to make an agreement. From a sequence of offering and rejecting, they could infer the relationship values of the other side. This process is similar to a two-player bargaining game in which rational players will reach an agreement with $\frac{2}{3} > \alpha > \frac{1}{3}$ in the bargaining.

How players choose relationships to maximize payoffs can be a complex problem because it depends on their knowledge, experience, and strategic interactions in the game. Consider Column player in the above game as an example. Column player would set $R_{CR}$ to the value so that $\frac{1}{1-{}_C\overline{R}_{RC}} = \frac{-R_{CR}}{1-R_{CR}}$, in which case Column player's payoff would be maximized if an agreement could be made. Given that ${}_C\overline{R}_{RC} = -0.5$, there will be $R_{CR} = -2$. Column player would reject any offer with $\alpha < \frac{2}{3}$ hoping that Row player would offer $\alpha \geq \frac{2}{3}$ in the next round. On the other hand, Row player would choose $R_{RC} = -2$ and would not offer any $\alpha \geq \frac{1}{3}$. As a consequence, no agreement could be reach. Realized that consequence, two players have to make concessions by choosing reasonable values of relationships, i.e. $0.5 > R_{CR} > -2$ and $0.5 > R_{RC} > -2$.

## 6. Conclusions

A generalized model of games that takes into consideration the relationships between players is proposed. Cooperative games and non-cooperative games are the special case of this model. There exists a significant set of games that are neither cooperative nor non-cooperative, which have not been investigated in game theory before.

We prove that there must exist a Nash equilibrium for every game under the relationship model so that equilibrium analysis developed in non-cooperative game theory can be applied. A prisoner's dilemma and a repeated ultimatum game are analyzed as applications of the relationship model.

One advantage of the relationship model lies in that it provides an accurate description of the players' attitudes toward others in game playing. Apart from cooperative and non-cooperative, a player's attitude toward another player could be sub-cooperative, hostile and dedicated. By taking supposed relationships into account, the attitudes of altruism and exploitation can be considered. Relationships and relationship changes make the strategies of players interdependent. Our future work will address the challenging issue of how the players take advantage of the relationships and the corresponding relationship changes.